# Optical Self-Trapping and Nonlinear Light–Matter Interactions in Biological Soft Matter

Lu Tian[1,#], Xianyang Liang[1,#], Liqin Tang[2], Rekha Gautam[3], Anna Bezryadina[4], Yu-Xuan Ren[5,*], Yi Liang[1,*], Zhigang Chen[2,*]

[1]*Guangxi Key Lab for Relativistic Astrophysics, Center on Nanoenergy Research, School of Physical Science and Technology, Guangxi University, Nanning, Guangxi 530004, China*

[2] *The MOE Key Laboratory of Weak-Light Nonlinear Photonics, School of Physics and TEDA Applied Physics Institute, Nankai University, Tianjin 300457, China*

[3]*Biophotonics@Tyndall, IPIC, Tyndall National Institute, Lee Maltings Complex, Cork, Ireland*

[4] *Department of Physics and Astronomy, California State University Northridge, Northridge CA 91330, USA*

[5]*Institute for Translational Brain Research, Jinshan Hospital, Fudan University, Shanghai, 200032, China*

# *These authors contributed equally to this work.*

*Corresponding Authors: yxren@fudan.edu.cn; liangyi@gxu.edu.cn; biglasers@gmail.com*

**Abstract:**

Low-scattering, deep-penetration light transport in biological media remains a pivotal challenge for biophotonic technologies, including biomedical imaging, optical diagnostics, and photodynamic therapy. This review builds upon and extends our earlier studies of nonlinear optical self-trapping and optically induced waveguiding in biological suspensions, such as human erythrocytes and cyanobacteria, where light–matter coupling is governed by optical-force-mediated particle redistribution. Recent progress has revealed increasingly rich and complex regimes, including the propagation and nonlinear self-action of structured (vortex) beams in biological environments, as well as nonlinear responses dominated by thermally driven mechanisms in absorptive biomolecular solutions (e.g., heme and chlorophyll). We place particular emphasis on distinctive nonlinear phenomena observed in these systems, including spatial self-phase modulation, optical-force-induced sculpturing of effective energy landscapes, and quasi-waveguide formation in soft, heterogeneous biological media. We conclude by highlighting emerging opportunities to harness these nonlinear behaviors for deep-tissue imaging, label-free biosensing, and the realization of biocompatible photonic structures and devices assembled directly from living or hybrid biological matter.

# 1. Introduction

Biological materials span a broad range of length scales and are commonly classified as soft matter, characterized by dynamic geometries and pronounced responsiveness to external stimuli such as heat, mechanical stress, and electromagnetic fields. As a result, biological media provide a particularly rich and versatile platform for nonlinear photonics [1-3]. In contrast to solid-state materials, biological systems often exhibit intrinsic heterogeneity, strong dissipation, and structural adaptability [4, 5], leading to optical responses that arise from multiple mechanisms acting concurrently—including optical forces, absorption, thermal effects, and stimulus-induced morphological changes in cells or biomolecules [6-10].

Early studies demonstrated that optical forces can spatially reorganize cells in colloidal suspensions containing micron-scale scatterers such as red blood cells, cyanobacteria, and *Escherichia coli* [2, 6, 7, 11, 12]. The resulting redistribution of particles along the beam path modifies the local refractive index landscape, enabling the formation of high-index channels. This process gives rise to optically induced self-focusing and waveguiding, which effectively mitigate scattering losses and substantially enhance light penetration through turbid biological media.

When a laser beam propagates through a biological suspension, a wide variety of behaviors emerge. At low input powers, photons undergo diffusive transport, and strong linear scattering dominates the propagation dynamics, resulting in output beam profiles that exhibit speckle-like intensity patterns. Such speckle patterns are spatially incoherent and typically detrimental for imaging, trapping, or communication. Nevertheless, their impact can be mitigated using advanced techniques, such as wavefront shaping, time-resolved speckle contrast imaging, and other computational approaches that recover high-fidelity images through multiple-scattering media [13-16]. Moreover, these dynamically evolving speckle fields offer a compelling testbed for studying optical rogue-wave phenomena, as interactions between light and fluctuating biological environments (e.g., Brownian motion of cells) generate nontrivial intensity statistics [17].

As the input power increases, nonlinear effects progressively emerge and counteract scattering. In cellular suspensions, optically induced forces can rearrange micron-scale particles and form stable, self-induced waveguide channels. In contrast, in biomolecular solutions containing nano-scale particles — such as heme, chlorophyll, or other strongly absorbing dyes — shifts toward photothermal nonlinearities [9, 18]. Here, absorption-induced heating modifies the refractive index via the thermo-optic effect. Under sufficiently high optical power, temperature gradients may further drive convective flows, producing corrugated beam propagation patterns that can even become gravity dependent [19].

Recent progress reveals two major developmental trends. First, the field has expanded from cellular suspensions to nanoscale molecular solutions. In systems containing biomolecules such as heme [9], chlorophyll [18, 20], and food dyes [8, 21], or plasmonic nanoparticles supporting localized surface plasmon resonances (LSPR) [22-27], the nonlinear optical response becomes significantly enhanced and increasingly multifaceted. Photothermal refractive changes, fluorescence emission, and intermolecular energy transfer may coexist and interact [18, 20, 21]. Remarkably, even in regimes typically dominated by self-defocusing, self-collimated beam propagation can be achieved under carefully engineered conditions [21]. Second, the research frontier has expanded from Gaussian beams to structured light fields, including vortex beams carrying orbital angular momentum (OAM), which play essential roles in micromanipulation and optical communication [28, 29]. Investigations of OAM preservation, nonlinear phase accumulation, and topological-charge conservation in biological media—each governed by distinct physical mechanisms—constitute a crucial step toward the realization of biocompatible photonic devices and light-based diagnostics. An overview of the nonlinear response landscape of bio–soft-matter systems is summarized in Fig. 1.

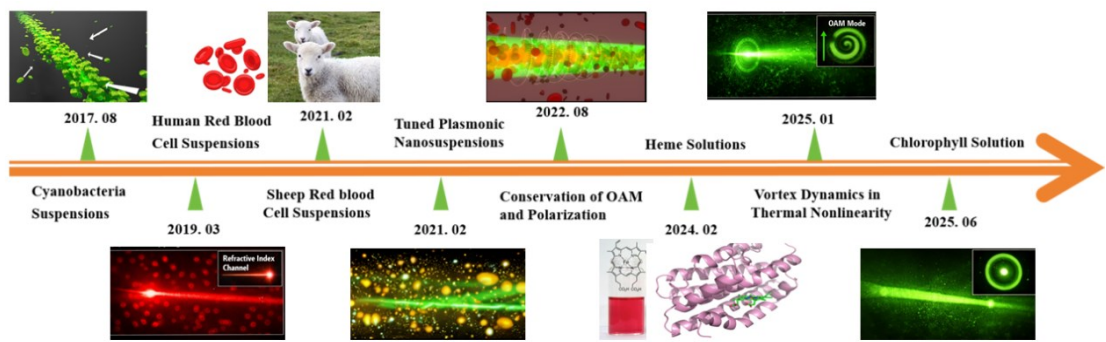

**Fig 1.** Snapshots of exemplary demonstrations of nonlinear beam propagation in bio-soft matter since the first nonlinear self-trapping experiment in 2017 [6].

In this short review, we present a concise overview of recent progress in optical nonlinear effects in biological soft matter and outline emerging research directions. We begin by revisiting key physical mechanisms, including optical-force-mediated self-focusing and thermally driven self-defocusing nonlinearities. We then highlight representative recent advances, such as OAM conservation in blood-cell suspensions, self-collimated light transmission in heme- and dye-based solutions, and fluorescence-assisted beam guidance and topological-charge detection in chlorophyll systems. Collectively, these studies underscore new opportunities for deep-tissue imaging, noninvasive biosensing, and the realization of photonic functionalities implemented directly within biological soft matter.

## 2. Principles of nonlinear responses in biological soft matter

The self-focusing nonlinearity driven by optical gradient and scattering forces in colloidal and soft-matter systems was well documented in our previous review [2]. Here we briefly summarize the essential mechanism. Optical forces are critical in the nonlinear optical (NLO) behavior of such media. For a small dielectric (Rayleigh) particle in an optical field (E), the induced polarization can be expressed as [31, 32]:

$$p = 4\pi\varepsilon_0 n_b^2 \left(\frac{m^2-1}{m^2+2}\right) r^3 E \equiv \alpha E, \quad (1)$$

where relative refractive index $m = n_p/n_b$, $n_p$ and $n_b$ are the refractive indices of the particle and background medium, respectively, $r$ is the particle radius, and $\alpha$ is the particle polarizability. A particle in an optical field experiences a gradient force ($F_g = \alpha\nabla I/4$) and a scattering force ($F_{sc} = I\sigma n_b/c$) [33-35], where $I$ is the optical intensity, $\sigma = \frac{128\pi^5 r^6 n_b^4}{3\lambda^4}\left(\frac{m^2-1}{m^2+2}\right)^2$ is the scattering cross-section, $\lambda$ is the wavelength of the laser, and $c$ is the speed of light in vacuum. High-index particles ($n_p > n_b$) are attracted to the beam center, while low-index particles ($n_p < n_b$) are repelled, and both results in a local refractive index increase at the beam center and self-focusing (Fig. 2(a-b)). The effective nonlinear refractive index change can be expressed as [36]：

$$\Delta n_{NLF} = (n_p - n_b) V_p \rho_0 \left(e^{\frac{\alpha I}{4k_B T}} - 1\right), \quad (2)$$

where $V_p$ is the particle volume, $\rho_0$ is the initial density, $k_B$ is the Boltzmann constant, and $T$ is the temperature. In this case, the beam propagation can be described by a nonlocal nonlinear Schrödinger equation:

$$i\frac{\partial}{\partial z}\varphi + \frac{1}{2kn_b}\nabla_\perp^2 \varphi + k\Delta n_{NLF}\varphi + i\gamma\varphi = 0 \quad (3)$$

where $k = 2\pi/\lambda$ is the wavevector in the media, $\varphi$ is the field envelope, and $\gamma = \sigma\rho_I/2$ is defined as the loss coefficient, which is proportional to the particle number density, i.e., the scattering loss depends on the beam intensity.

In soft-matter systems, absorption-induced thermal effects become significant, introducing refractive index gradients (thermal lensing) and thermophoresis, where particles migrate along temperature gradients. When a Gaussian beam propagates through a strongly absorbing solution, absorption converts photon energy into heat and local temperature rise that dictates the refractive index profile via the thermo-optic effect. Under gravity, these gradients can further drive heat diffusion and convection, leading to asymmetric thermal lenses and spatio-temporal beam distortion. In the long-time limit, where thermal diffusion dominates over the finite beam waist and neglect the finite beam-waist effects ($8Dt \gg \omega^2$), the resulting temperature change $\Delta T(x, y, t)$ inside the medium can be described as [37]:

$$\Delta T(x,y,t) = \frac{\alpha_0 P}{\pi \rho c_p} \left\{ \int_0^t \frac{dt'}{8Dt'} exp\left[-\frac{2[(y-v_y t')^2 + x^2]}{8Dt' + \omega^2}\right] \right\}, \quad (4)$$

where $c_p$ is the isobaric specific heat capacity, $P$ is the incident optical power, $\rho$ denotes the particle number density, $v_y$ is the convective velocity, $\alpha_0$ is the linear absorption coefficient, $\omega$ is the beam waist, and $D$ is the thermal diffusion coefficient. The refractive index change due to photothermal effect is given by [9]:

$$\Delta n_{NLT} = \frac{dn}{dT} \Delta T, \quad (5)$$

where $dn/dT$ is the thermo-optic coefficient with *dn/dT<0* yielding self-defocusing and *dn/dT>0* yielding self-focusing nonlinearity of suspension. In this case, the beam propagation can be described by a nonlocal nonlinear Schrödinger equation [36]:

$$i\frac{\partial}{\partial z}\varphi + \frac{1}{2k}\nabla_\perp^2 \varphi + k\frac{\Delta n_{NLT}}{n_b}\varphi + i\gamma\varphi = 0 \quad (6)$$

where *k = 2π/λ* is the wavevector in the media, $n_b$ is the refractive index of the background medium, $\varphi$ is the field envelope, ρ is the particle density, and γ is the loss coefficient.

In most liquids, the thermo-optic coefficient is negative, resulting in a refractive index decrease at the beam center and enhanced diffraction. However, by carefully positioning the beam focus within the absorbing medium, the thermal lensing can shift the effective focal point forward as input power increases [38]. Within a well-defined power window, this can produce a self-collimated propagation channel analogous to self-focusing and may even counteract diffraction to form optical spatial solitons, enabling stable, diffraction-free propagation, as shown in Fig. 2 (right).

Overall, optical-force-mediated and thermally-induced nonlinearities provide a versatile toolbox for achieving light self-trapping, beam guidance, and optical solitons in suspensions of inorganic nanoparticles or biocompatible soft matter, with potential applications in biomedical imaging, communication, and photonic devices.

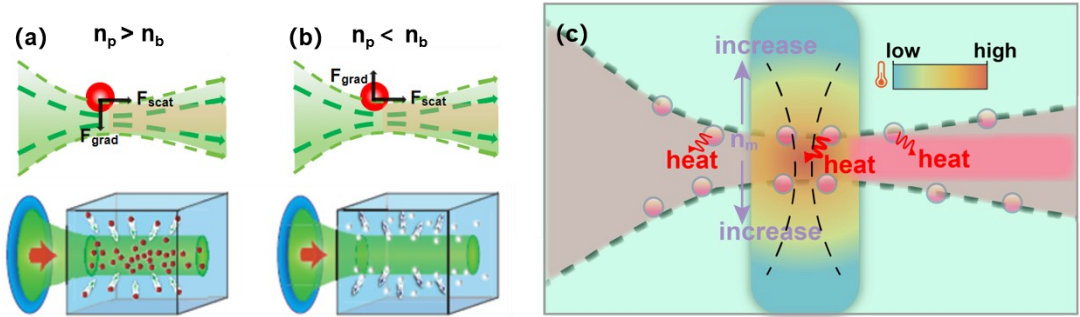

**Fig 2. Schematic illustrations of Gaussian beam propagation through nonlinear media.** (a) The formation of a self-collimated beam driven by optical gradient force and scattering force, where high-refractive-index particles are attracted toward the beam center while (b) low-refractive-index particles are repelled from it. (c) The formation of a self-collimated beam driven by the photo-thermal effect. Bottom panels in Figs 2. (a) - (b) are reproduced with permission from reference [2].

## 3. Biological soft matter as a nonlinear waveguiding platform

In the past several years, research on nonlinear light propagation in biological soft matter has evolved from optical-force-mediated self-action in cell suspensions to photothermal nonlinearities in absorptive molecular solutions [Fig. 1] and angular momentum preservation. This evolution has enabled a significant reduction in operating power, making nonlinear beam self-action compatible with biological suspensions, while expanding the range of achievable beam dynamics from self-guided Gaussian beams to structured-light propagation [6, 7, 9, 18, 30, 39].

Early efforts in the direction on nonlinear dynamics in biological soft matter focused on developing nonlinear medium using cell suspensions for waveguiding. In 2017, Bezryadina et al. first demonstrated nonlinear self-trapping of light in a highly scattering cyanobacteria suspension in seawater [6]. They showed that, in addition to the optical gradient force, forward-scattering forces acting on the cells play a decisive role in shaping the beam propagation. This conclusion was supported by both experiment and theoretical modeling. The incorporation of forward-scattering forces effectively resolved the beam-collapse issue that had appeared in the earlier 2013 experiments on dielectric-polymer-sphere suspensions [32]. Although the required power was relatively high (~3 W), this work laid a solid theoretical foundation for subsequent studies on cell-based waveguiding [Fig. 3(a)].

In 2019, Gautam et al. reported nonlinear self-trapping of light in human red blood cell (RBC) suspensions at substantially lower power levels (~200 mW) [7]. The RBCs, dispersed in phosphate-buffered saline, were studied under isotonic, hypotonic, and hypertonic conditions, enabling controlled modulation of the nonlinear response. At higher powers, they also observed the onset of thermal convection, marking the first exploration of thermo-fluidic effects in biological nonlinear waveguiding [Fig. 3(b)].

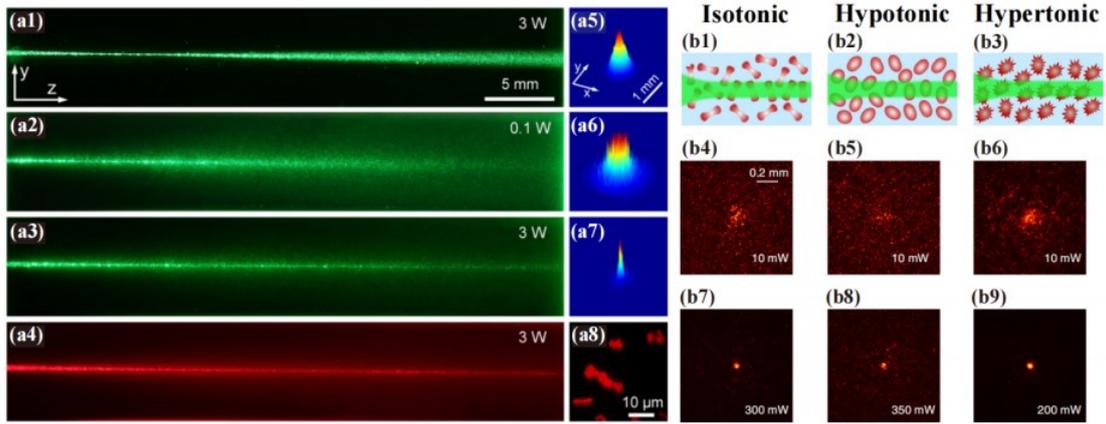

**Fig 3. Light propagation in cells suspensions over several centimeters.** (a) Nonlinear self-trapping of light through cyanobacteria in seawater. At low power, beam undergoes linear diffraction and scattering, while at high power, it experiences nonlinear self-trapping. (b) Self-trapping of light through human RBC suspensions under different osmotic conditions (isotonic, hypotonic, and hypertonic). Reproduced with permission from reference [6, 7, 27].

A major advance in biomolecular solutions occurred in 2024, when Zhang et al. demonstrated that heme solutions possess strong photothermal nonlinearity [9], lowering the optimal self-trapping power to ~70 mW. In contrast to cell suspensions, where optical forces drive particle redistribution, heme molecules primarily induce refractive-index gradients through absorption-mediated thermal effects. These gradients extend the beam's focus forward, enabling self-collimated propagation even in a self-defocusing medium. The output beam first narrows and then broadens with increasing input power [Fig. 4]. At still higher powers, thermal convection dominates, producing asymmetric corrugation patterns, while side-view images reveal gravity-induced beam bending [Fig. 4(e)].

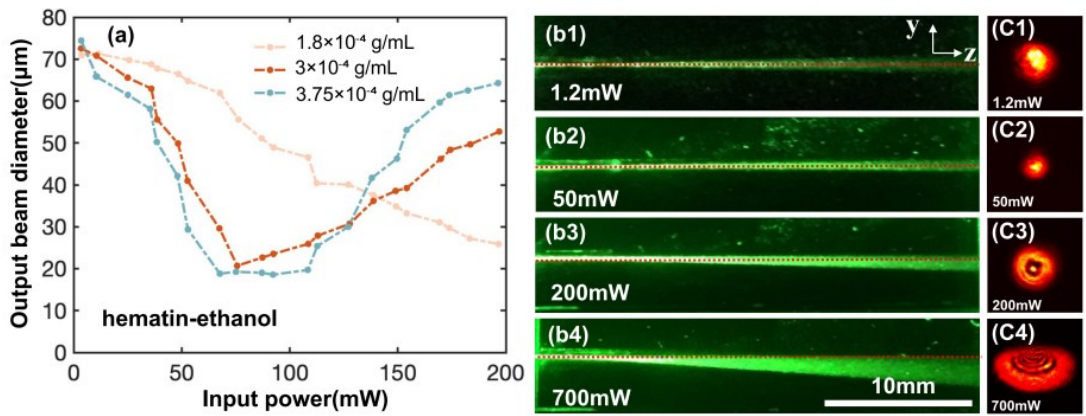

**Fig 4. Photothermal nonlinearity in heme molecule suspensions.** (a) Measured output beam diameter as a function of optical power in hematin-ethanol solutions. (b1-b4) Side-view profiles of beam propagation at different optical powers in hematin-ethanol solutions with mass concentration of $3.75\times10^{-4}$ g/mL. (c1-c4) Output patterns corresponding to the same power used in (b1-b4). Reproduced with permission from reference [9].

In 2025, the required power for self-trapping was dramatically reduced to the sub-10 mW regime in natural chlorophyll solution [18], a biologically abundant photosensitive molecule [Fig. 5]. An exponential-decay model was employed to systematically clarify the principles underlying nonlinear beam propagation in such absorptive molecular solution. Chlorophyll concentration not only tunes the optimal self-trapping power and output beam size but also defines the stable operating range for the biological waveguide-like propagation [Figs. 5(a-c)]. Lower concentrations require higher power for waveguide formation but simultaneously reduce thermal load, thereby mitigating photodamage and enabling stable self-guided propagation over a wider power range. Additionally, the chlorophyll fluorescence excited by a 532-nm pump closely follows an exponential decay and exhibits a propagation mode identical to that of the pump laser [Figs. 5(e-j)].

The structured beams carry information, and transmit through scattering media. In particular, vortex beams, characterized by helical wavefronts and tunable orbital angular momentum (OAM), are increasingly important for quantum information and high-dimensional optical communication [40]. Pérez et al. conducted a pioneering study of vortex-beam propagation in sheep RBC suspensions [30]. In buffer solution alone, vortex beams at 532 nm or 780 nm experience only linear diffraction while maintaining phase singularity. In self-focusing RBC suspensions, despite strong scattering and nonlinear focusing, the topological charge remains unchanged across several centimeters of propagation [Figs. 6(a-d)]. The OAM is therefore conserved over several centimeters of interaction with the medium, which does not scramble the angular-momentum state. Pump–probe experiments further revealed that bio-waveguides generated under three scenarios, including Gaussian-guiding-vortex, vortex-guiding-vortex, and vortex-guiding-Gaussian, all effectively coupled low-power probe beams while preserving OAM [Figs. 6(e-j)], indicating the formation of a localized high-index core inside the suspension. The output polarization (linear or circular) is similarly preserved for pump and probe beams through scattering media [Figs. 6(k-l)]. Theoretical analysis shows that self-formed light channels in biological media limit scattering, allowing twisted laser beams to propagate over long distances without altering the information. Further studies shown that higher-order OAM beams is nearly preserved during propagation through complex turbid media, although scattering-induced phase distortions and mode mixing are unavoidable [39].

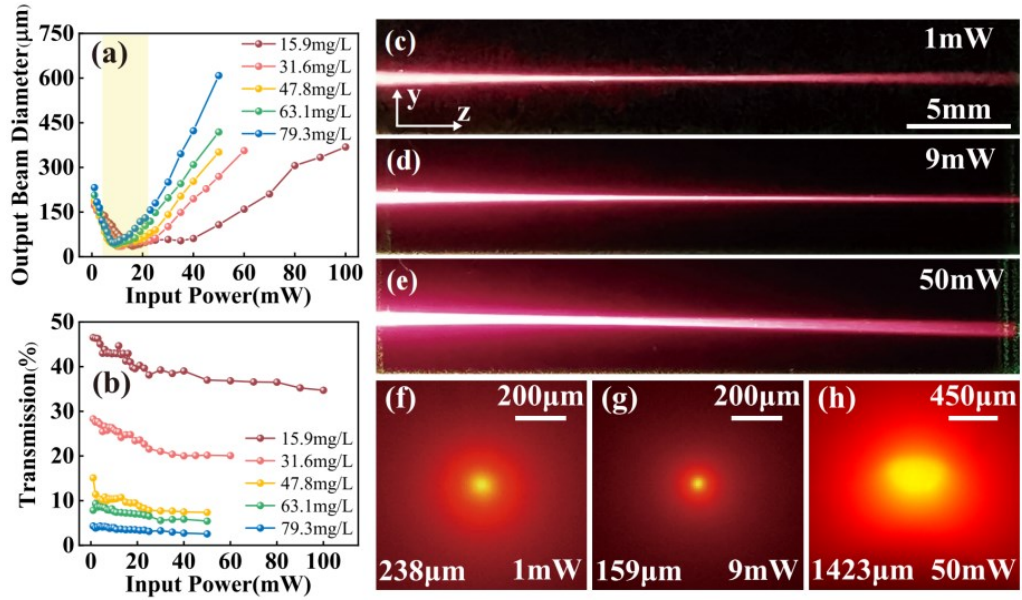

**Fig 5. Concentration-dependent optical nonlinearity and the propagation of fluorescence in chlorophyll solution.** (a) The output diameter and (b) transmission as function of the concentration of chlorophyll. (c-e) Side-views and (f-h) output intensity profiles of chlorophyll fluorescence induced by 532nm laser at respective power. Reproduced with permission from reference [18].

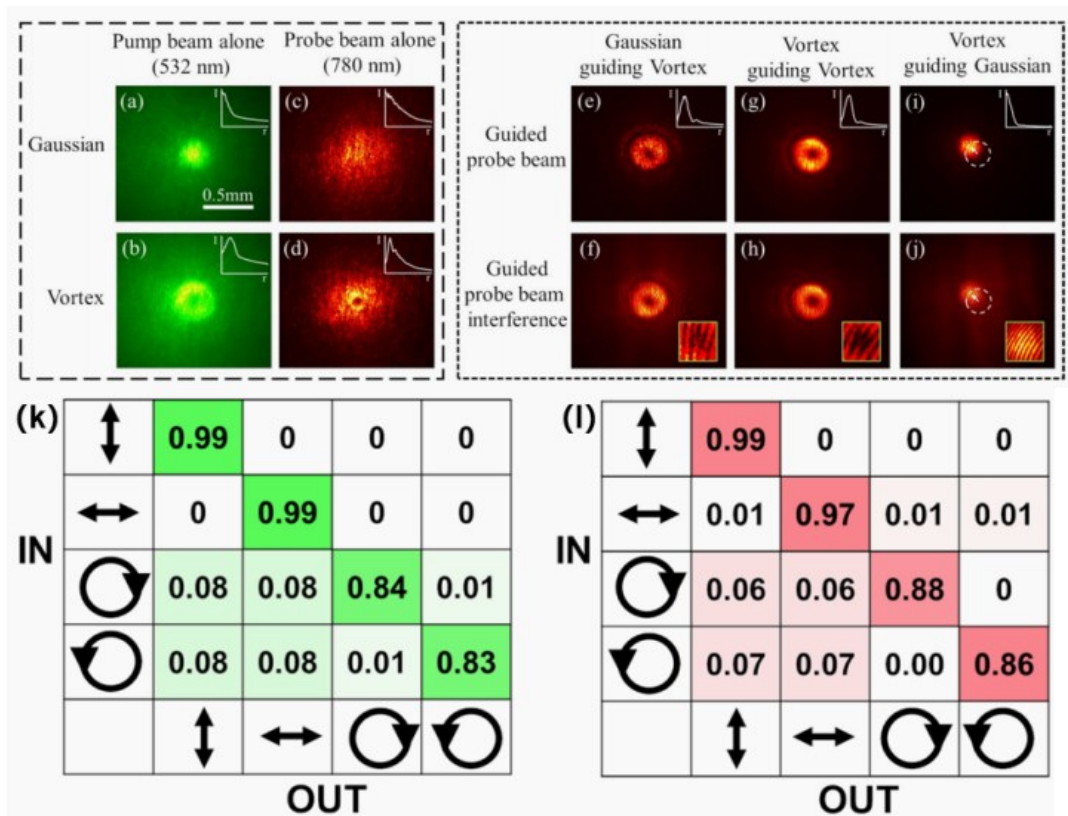

**Fig 6. Nonlinear coupling of a NIR (780 nm) beam and a green (532 nm) beam through a sheep RBC suspension.** (a, b) Gaussian and vortex green pump beam profiles at the output with optimal power for waveguide formation (500 mW). (c, d) Gaussian and vortex NIR probe beam profiles at a weak power of 20 mW. (e) Guiding of the vortex beam by the green Gaussian beam. (g) Guiding of the vortex beam by the green vortex beam. (i) Guiding of the Gaussian NIR beam by the green

vortex beam. The probe beam shifts to one side of the vortex beam with the greatest intensity. A dashed circle shows the unguided position of the Gaussian beam. (f, h, j) Interferogram of the guided probe beams (e, g, i), respectively. All images are beam profiles at the exit face of the 3-cm cuvette through the sheep RBC suspension. The output intensity distribution between polarization states (vertical, horizontal, right circular, and left circular) after propagating through the RBC suspension for different input polarization states of the (k) pump beam alone (532 nm) at 500 mW and (l) the probe beam (780 nm) with guiding. Reproduced with permission from reference [30].

Complementing these developments, the nonlinear dynamics of vortex beams with different topological charges were recently investigated in a strongly absorptive, thermally self-defocusing chlorophyll solution to understand how thermal nonlinearities govern structured-light transport in biological media [20]. At optimized powers, a fundamental vortex ($l = 1$) forms a quasi-optical waveguide, whereas higher-order vortices ($|l| > 1$) undergo thermally driven azimuthal instabilities leading to charge-dependent core fragmentation and asymmetric far-field patterns [Fig. 7]. These power-dependent spatial self-phase modulation effects reveal the microdynamics of the optofluidics and provide a simple, passive route for identifying vortex topological charge, highlighting the potential of thermally nonlinear media for biophotonic sensing.

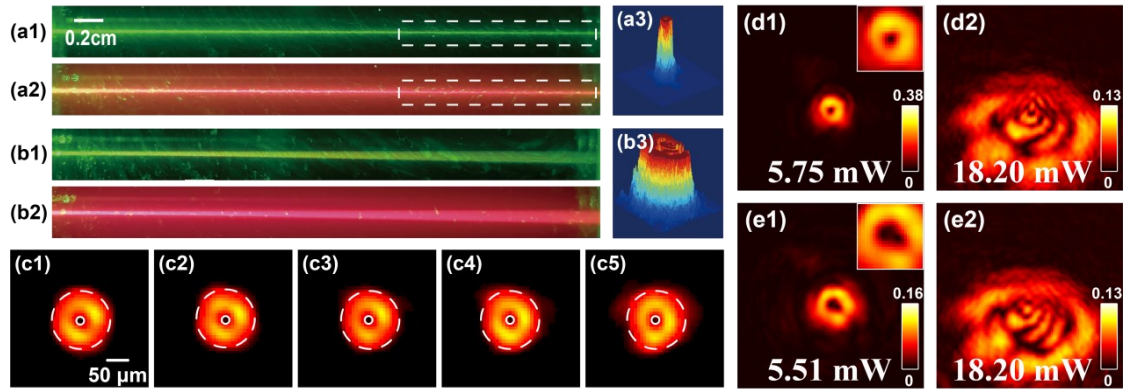

**Fig 7. Propagation of vortex beams in chlorophyll solution.** (a1) The side view of the fundamental vortex beam passing through the chlorophyll solution, with fluorescence filtered out, under moderate and (b1) high power, respectively. (a2) and (b2) are corresponding side views with fluorescence. (a3) The 3D intensity profile of the output plane of the fundamental vortex beam under moderate and (b3) high power, respectively. (c1)–(c5) The cross-sectional profiles of the fundamental vortex beam propagating through the chlorophyll solution at moderate power, captured at distances of 0 cm, 0.3 cm, 0.6 cm, 0.9 cm, and 1.2 cm from the output plane, corresponding to the waveguide section formed within the dashed box in (a1)–(a2). The splitting of singularities in the output plane for vortex beams with topological charges of (d1) l = 1 and and (e1) l = 2, respectively, under moderate power. The formation of asymmetric thermal rings and fork-shaped fringes in the output plane for vortex beams with (d2) l = 1 and (e2) l = 2, respectively, under high power. Reproduced with permission from reference [20].

## 4. Summary and Outlook

Light propagation in biological soft-matter systems, including biomolecular solutions, cellular suspensions, and living tissues, has attracted increasing interest across both fundamental and application-driven research. Despite substantial progress, many nonlinear and wave-transport phenomena in these complex media remain only partially understood. Recent observations in metallic nanoparticle suspensions, for example, have revealed spin-locking behavior of optical beams [41], suggesting that biological environments may host similarly rich and unconventional light–matter interactions. Unlike engineered materials, natural biological particles exhibit inhomogeneous composition, irregular morphology, and intrinsic symmetry breaking, giving rise to highly complex optical landscapes. These features can induce optical responses that are difficult to resolve or entirely absent in conventional systems, underscoring the need for deeper exploration of nonlinear, thermal, and structural effects in biological media.

Meanwhile, achieving tight light focusing and controlled guidance in strongly scattering environments remains a central challenge for non-invasive imaging, diagnostics, and sensing, particularly within the visible and near-infrared spectral windows relevant to biophotonics. Notably, preliminary studies indicate that gold-nanorod colloidal suspensions can enhance the fluorescence of trace concentrations of Rhodamine 6G (0.2 nM) by more than fifteenfold through nonlinear self-action of a 532-nm probe beam [27]. Similar enhancement has been observed in low-concentration *E. coli* suspensions through plasmonic-resonant, self-guided propagation. In these systems, fluorescence amplification emerges from the interplay of plasmonic resonance, nonlinear refractive-index modification, and self-induced light channels. Such mechanisms provide promising routes for detecting weak biological signals, improving imaging contrast, and enabling compact, multiplexed optical sensing architectures. More recently, advances in adaptive wavefront control have begun to address light delivery through dynamic and highly scattering biological media. In particular, absorption-guided time-reversal approaches exploit intrinsic nonlinear feedback to achieve deep, speckle-scale optical focusing without the need for implanted or point-like guide stars [16].

With the rapid development of artificial intelligence, optical science has been significantly accelerated by data-driven approaches. Techniques such as inverse design [42–44], asymmetric training [45], and backpropagation-based optical neural networks [46] have enabled rapid convergence toward target optical fields. These developments are reshaping established paradigms in structured-light control [47], deep non-invasive focusing [13], and speckle imaging [48–50], while substantially enhancing the robustness and generality of optical systems [51,52]. A particularly promising direction is the application of optical neural networks to wavefront shaping in biological soft matter. Such approaches are expected to enable multi-beam structured-light control, increase the dimensionality of optical manipulation, and mitigate thermally induced convective effects at high optical powers. More broadly, interdisciplinary integration of nonlinear photonics, soft-matter physics, and machine-learning-assisted control is anticipated to facilitate rapid identification and real-time detection of early-stage diseases [53], as well as to promote efficient optical multiplexing [54,55] and high-capacity information encoding in optical communication systems [56,57].

Collectively, these advances point to a rich frontier at the intersection of nonlinear photonics, optical forces, soft-matter physics, and AI-empowered biophotonics [58-61]. Future efforts will likely focus on integrating structured light, plasmonic enhancement, and adaptive wavefront control to overcome scattering, extend penetration depth, and exploit the intrinsic nonlinearities of biological materials. Looking ahead, the combination of data-driven control strategies with nonlinear biophotonics is expected to enable real-time optimization of structured-light transport in fluctuating biological environments, opening pathways toward intelligent photonic tools for deep-tissue analysis and biosensing (Fig. 8).

This review has briefly surveyed recent progress in nonlinear light propagation in biological soft-matter systems, spanning micron-scale cellular suspensions and nanoscale biomolecular solutions, and encompassing both Gaussian and structured light excitation. In nanoparticle and cellular suspensions, optical forces drive particle redistribution and the formation of self-induced refractive-index channels, enabling waveguiding and enhanced transmission through strongly scattering media. In contrast, biomolecular and nanoscale systems are governed primarily by photothermal nonlinearities, which give rise to quasi-waveguide behavior, spatial mode reshaping, and fluorescence-assisted beam control. The preservation of orbital angular momentum and polarization during nonlinear propagation points to the feasibility of high-dimensional light transport in complex biological environments, while thermally driven effects enable functionalities such as topological-charge detection and adaptive beam guidance that are challenging to realize in conventional materials. Together, these findings establish biological soft matter as an active nonlinear optical medium, rather than a passive scattering background, and motivate continued exploration of its unique light–matter interactions for advanced biophotonic applications.

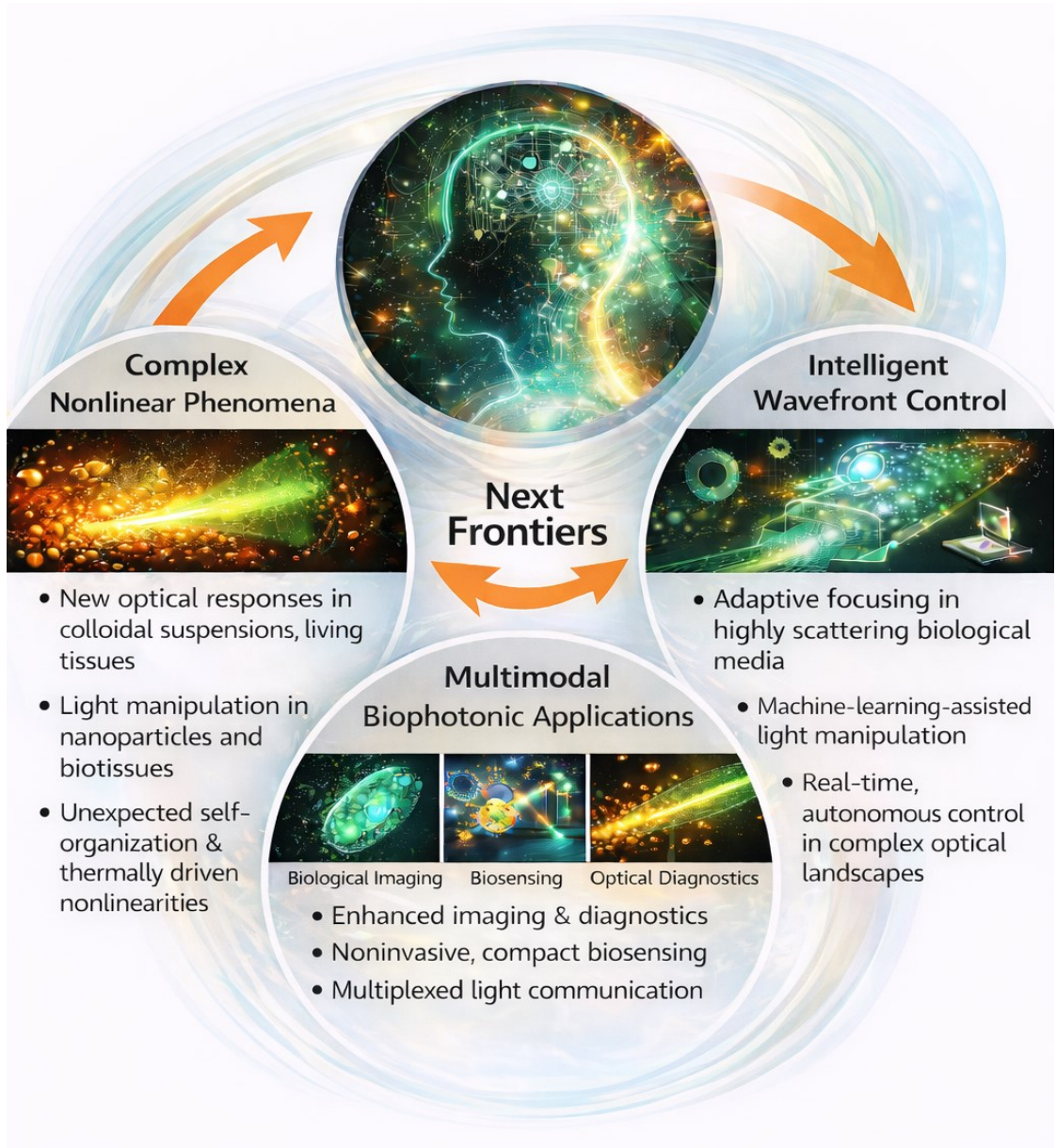

**Fig 8. Summary and outlook of nonlinear photonics in biological soft matter.** Schematic overview of emerging frontiers in nonlinear light–matter interactions in biological soft-matter systems, highlighting complex nonlinear phenomena in heterogeneous media, intelligent wavefront control enabled by adaptive and machine-learning-assisted approaches, and multimodal biophotonic applications. The figure illustrates how structured light, nonlinear self-action, and adaptive control may converge to enable controlled light transport, enhanced imaging, and biosensing in strongly scattering and dynamically evolving biological environments [created with Chat GPT].

**Acknowledgement:**

This work was supported by the National Natural Science Foundation of China (No. 12474290, W2541003, 12134006 and 62475047), by the National Key R&D Program of China (No. 2022YFA1404800), the Guangxi Natural Science Foundation (No. 2024GXNSFAA010314), and the special funding for Guangxi Bagui Youth Scholars (Yi Liang).